\documentclass[a4paper]{aa}
\usepackage[T1]{fontenc}
\usepackage[utf8]{inputenc}
\usepackage{arabtex}
\usepackage{utf8}
\usepackage{graphicx}
\usepackage{color}
\usepackage{txfonts}
\usepackage{microtype}
\usepackage{ellipsis}
\usepackage{natbib}

\bibpunct{(}{)}{;}{a}{}{,}
\usepackage[pdftex,colorlinks=true,urlcolor=blue]{hyperref}
\usepackage[xindy,nonumberlist,nopostdot,nogroupskip]{glossaries}

\makeglossaries
\usepackage[xindy]{imakeidx}
\makeindex

\begin{document} 
\title{Navigating with the Kamal}
\author{M. Nielbock}
\institute{Haus der Astronomie, Campus MPIA, Königstuhl 17, D-69117 
Heidelberg, Germany\\
\email{nielbock@hda-hd.de}}

\date{Received July 14, 2016; accepted }

\abstract{The students build and use an old navigational tool from the Arab world of the 9th century, the kamal. After an introduction to historic seafaring and navigation, they build this simple tool and understand how it can be used to measure angles. After learning that the elevation of Polaris is (almost) identical to the latitude of the observer, they apply this knowledge while using the kamal. During a field trip, they actually measure the elevation of Polaris. The result is compared with modern methods.}

\keywords{Earth, navigation, countries, astronomy,  history, geography, stars, Polaris, North Star, equator, latitude, longitude, meridian, pole height, celestial navigation, Arabia, kamal}

\maketitle
%

\section{Background information}

\subsection{Latitude and longitude}
\begin{figure}[!ht]
 \resizebox{\hsize}{!}{\includegraphics{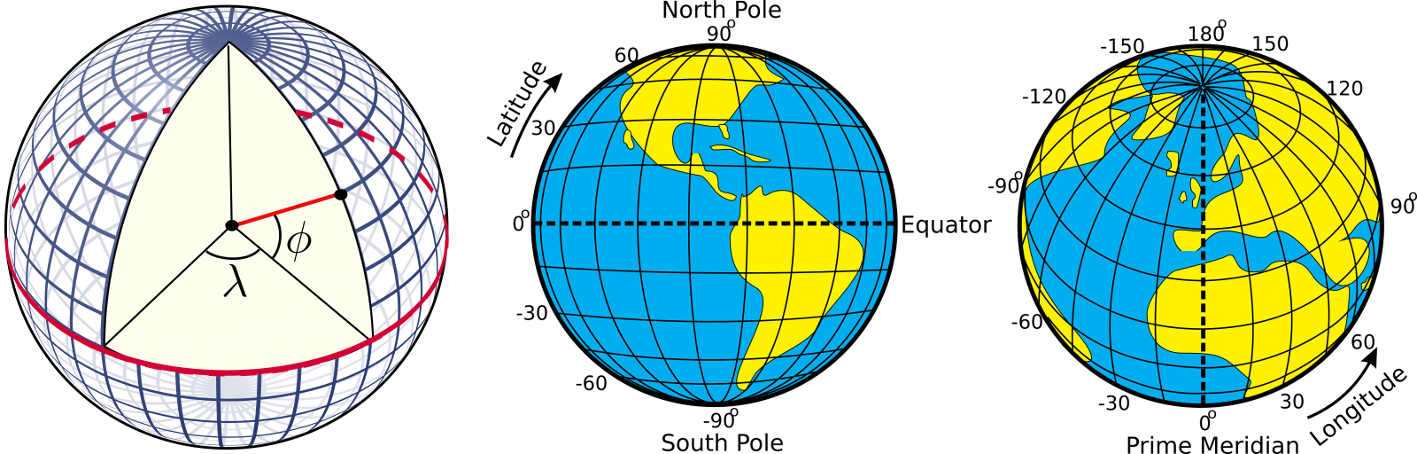}}
 \caption{Illustration of how the latitudes and longitudes of the Earth are 
defined (Peter Mercator, djexplo, CC0).}
 \label{f:latlong}
\end{figure}

Any location on an area is defined by two coordinates. The surface of a sphere 
is a curved area, but using coordinates like up and down does not make much 
sense, because the surface of a sphere has neither a beginning nor an ending. 
Instead, we can use
\newglossaryentry{spherical}
{
         name = {Spherical polar coordinates},
  description = {The natural coordinate system of a flat plane is Cartesian and measures distances in two perpendicular directions (ahead, back, left, right). For a sphere, this is not very useful, because it has neither beginning nor ending. Instead, the fixed point is the centre of the sphere. When projected outside from the central position, any point on the surface of the sphere can be determined by two angles with one of them being related to the symmetry axis. Such axis defines two poles. In addition, there is the radius that represents the third dimension of space, which permits determining each point within a sphere. This defines the spherical polar coordinates. When defining points on the surface of a sphere, the radius stays constant.}
}
spherical polar coordinates originating from the centre of the sphere with the radius being fixed (Fig.~\ref{f:latlong}). Two angular coordinates remain. Applied to the Earth, they are called the latitude and the longitude. Its rotation provides the symmetry axis. The North Pole is defined as 
the point, where the theoretical axis of rotation meets the surface of the sphere and the rotation is counter-clockwise when looking at the North Pole from above. The opposite point is the South Pole. The equator is defined as the great circle
\newglossaryentry{great}
{
         name = {Great circle},
  description = {A circle on a sphere, whose radius is identical to the radius of the sphere.}
}
half way between the two poles.

The latitudes are circles parallel to the equator. They are counted from 
$0\degr$ at the equator to $\pm 90\degr$ at the poles. The longitudes are great 
circles connecting the two poles of the Earth. For a given position on Earth, 
the longitude going through the
\newglossaryentry{zenit}
{
         name = {Zenith},
  description = {Point in the sky directly above.}
}
zenith, the point directly above, is called the meridian. This is the line the Sun apparently
\newglossaryentry{appa}
{
         name = {Apparent movement},
  description = {Movement of celestial objects in the sky which in fact is caused by the rotation of the Earth.}
}
crosses at local noon. The origin of this coordinate is defined as the
\newglossaryentry{meri}
{
         name = {Meridian},
  description = {A line that connects North and South at the horizon via the zenith.}
}
Prime Meridian, and passes Greenwich, where the Royal Observatory of England is located. From there, longitudes are counted from $0\degr$ to $+180\degr$ (eastward) and $-180\degr$ (westward). 

Example: Heidelberg in Germany is located at 49\fdg4 North and 8\fdg7 East.

\subsection{Elevation of the pole (pole height)}
\begin{figure}[!ht]
 \resizebox{\hsize}{!}{\includegraphics{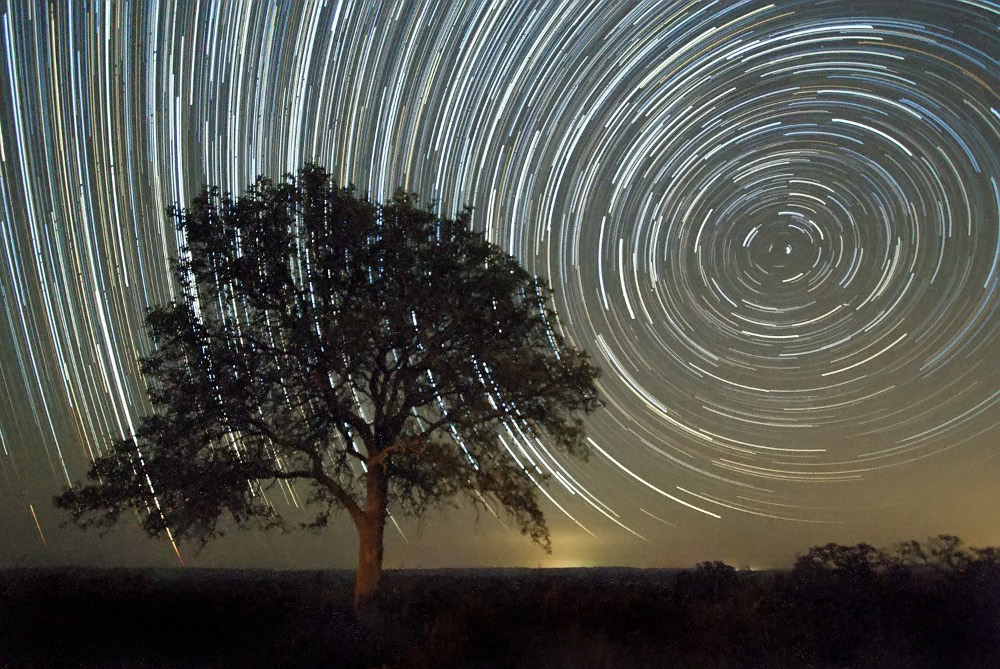}}
 \caption{Trails of stars at the sky after an exposure time of approximately 2 
hours (Ralph Arvesen, Live Oak star trails, 
\url{https://www.flickr.com/photos/rarvesen/9494908143}, 
\url{https://creativecommons.org/licenses/by/2.0/legalcode}).}
 \label{f:trails}
\end{figure}

If we project the terrestrial coordinate system of latitudes and longitudes at 
the sky, we get the celestial equatorial coordinate system. The Earth's equator 
becomes the celestial equator and the geographic poles are extrapolated to build 
the celestial poles. If we were to make a photograph with a long exposure of the 
northern sky, we would see from the trails of the stars that they all revolve 
about a common point, the northern celestial pole (Fig.~\ref{f:trails}).

\begin{figure}[!ht]
 \resizebox{\hsize}{!}{\includegraphics{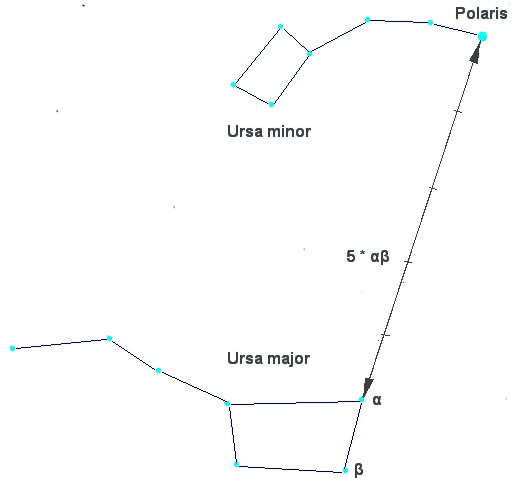}}
 \caption{Configuration of the two constellations Ursa Major (Great Bear) and Ursa Minor (Little Bear) at the northern sky. Polaris, the North Star, which is close to the true celestial north pole, is the brightest star in Ursa Minor (Bonč, \url{https://commons.wikimedia.org/wiki/File:Ursa_Major_-_Ursa_Minor_-_Polaris.jpg}, ``Ursa Major – Ursa Minor – Polaris'', colours inverted by Markus Nielbock, \url{https://creativecommons.org/licenses/by-sa/3.0/legalcode}).}
 \label{f:polaris}
\end{figure}

In the northern hemisphere, there is a moderately bright star near the celestial pole, the North Star or Polaris. It is the brightest star in the constellation of the Little Bear, Ursa Minor (Fig.~\ref{f:polaris}). In our era, Polaris is less than a degree off. However, 1000 years ago, it was $8\degr$ away from the pole. Therefore, today we can use it as a proxy for the position of the celestial north pole.

\begin{figure}[!ht]
 \centering
 \resizebox{0.45\hsize}{!}{\includegraphics{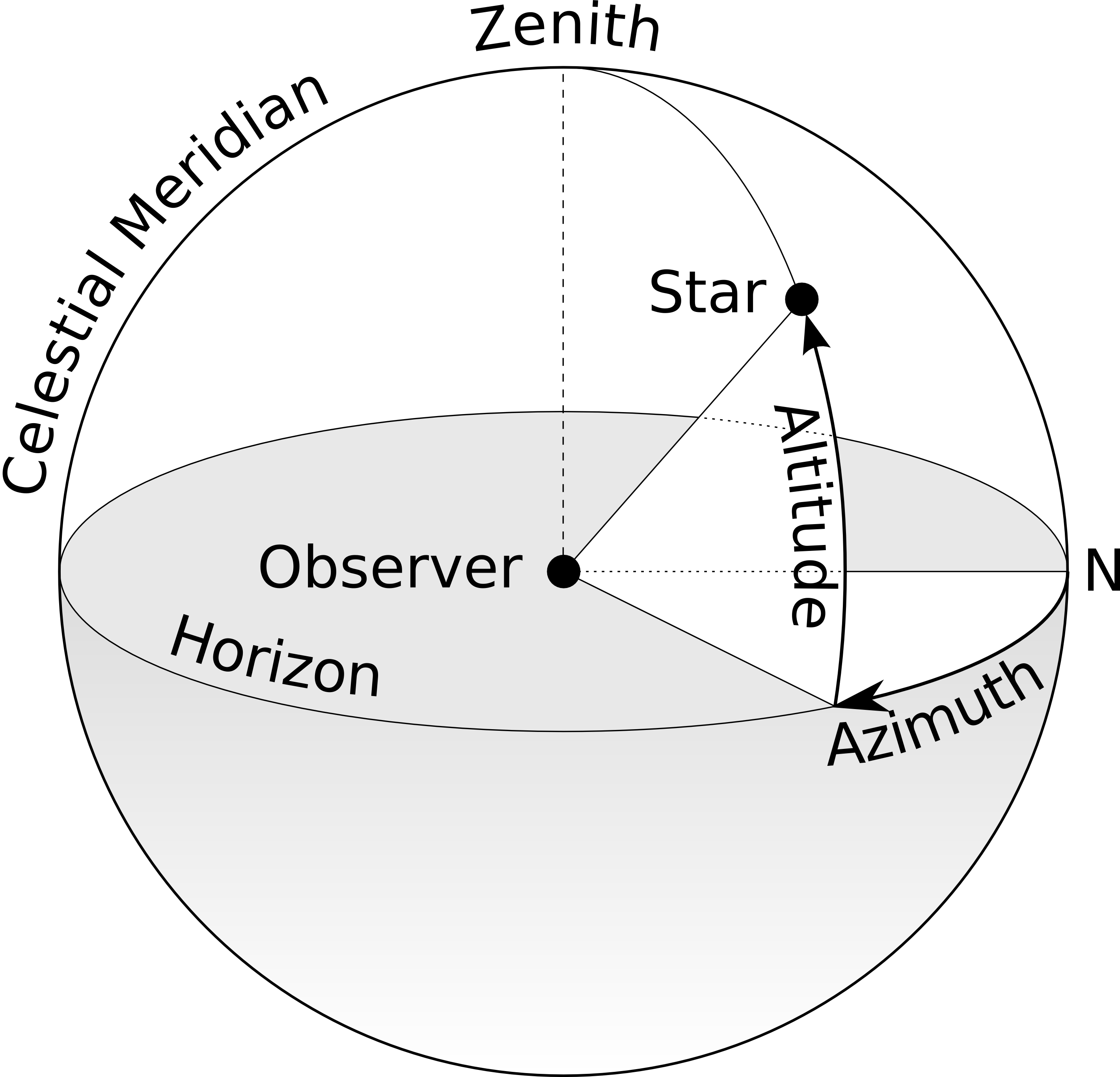}}
 \caption{Illustration of the horizontal coordinate system. The observer is the 
origin of the coordinates assigned as azimuth and altitude or elevation 
(TWCarlson, \url{
https://commons.wikimedia.org/wiki/File:Azimuth-Altitude_schematic.svg},
``Azimuth-Altitude 
schematic'', \url{https://creativecommons.org/licenses/by-sa/3.0/legalcode}).}
  \label{f:altaz}
\end{figure}

At the southern celestial pole, there is no such star that can be observed with the naked eye. Other procedures have to be applied to find it. At the southern celestial pole, there is no such star that can be observed with the naked eye. Other procedures have to be applied to find it. If we stood exactly at the geographic North Pole, Polaris would always be directly overhead. We can say that its elevation
\newglossaryentry{elev}
{
         name = {Elevation},
  description = {Angular distance between a celestial object and the horizon.}
}
would be (almost) $90\degr$. This information already introduces the horizontal coordinate system (Fig.~\ref{f:altaz}).

It is the natural reference we use every day. We, the observers, are the origin of that coordinate system located on a flat plane whose edge is the horizon. The sky is imagined as a hemisphere above. The angle between an object in the sky and the horizon is the altitude or elevation. The direction within the plane is given as an angle between $0\degr$ and $360\degr$, the azimuth, which is usually counted clockwise from north. In navigation, this is also called the bearing. The meridian is the line that connects North and South at the horizon and passes the zenith.

\begin{figure}[!ht]
 \resizebox{\hsize}{!}{\includegraphics{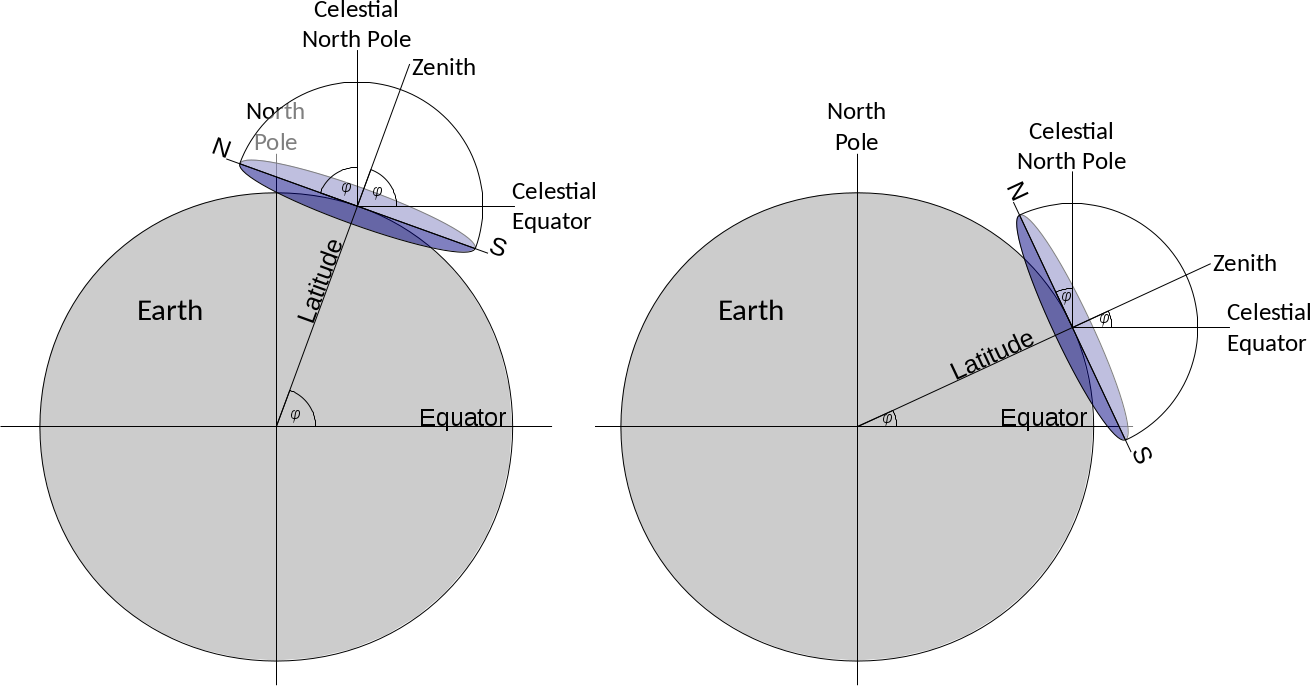}}
 \caption{When combining the three coordinate systems (terrestrial spherical, 
celestial equatorial, local horizontal), it becomes clear that the latitude of 
the observer is exactly the elevation of the celestial pole, also known as the 
pole height (own work).}
   \label{f:poleheight}
\end{figure}

For any other position on Earth, the celestial pole or Polaris would appear at 
an elevation smaller than $90\degr$. At the equator, it would just graze the 
horizon, i.e. be at an elevation of $0\degr$. The correlation between the 
latitude (North Pole = $90\degr$, Equator = $0\degr$) and the elevation of 
Polaris is no coincidence. Figure~\ref{f:poleheight} combines all three 
mentioned coordinate systems. For a given observer at any latitude on Earth, the 
local horizontal coordinate system touches the terrestrial spherical polar 
coordinate system at a single tangent point. The sketch demonstrates that the 
elevation of the celestial North Pole, also called the
\newglossaryentry{poleht}
{
         name = {Pole height},
  description = {Elevation of a celestial pole. Its value is identical to the latitude of the observer on Earth.}
}
pole height, is exactly the northern latitude of the observer on Earth. From this we can conclude that if we measure the elevation of Polaris, we can determine our latitude on Earth with reasonable precision.

\subsection{Triangles and trigonometry}
The concept of the kamal relies on the relations within triangles. Those are very simple geometric constructs that already the ancient Greeks have been working with. One basic rule is that the sum of all angles inside a triangle is $180\degr$ or $\pi$. This depends on whether the angles are measured in degrees or radians. One radian is defined as the angle that is subtended by an arc whose length is the same as the radius of the underlying circle. A full circle measures $360\degr$ or $2\pi$.

The sides of a triangle and its angles are connected via trigonometric functions, e.g.~sine, cosine and tangent. The easiest relations can be seen in right-angular triangles, where one of the angles is $90\degr$ or $\pi/2$.

\begin{figure}[!ht]
 \resizebox{\hsize}{!}{\includegraphics{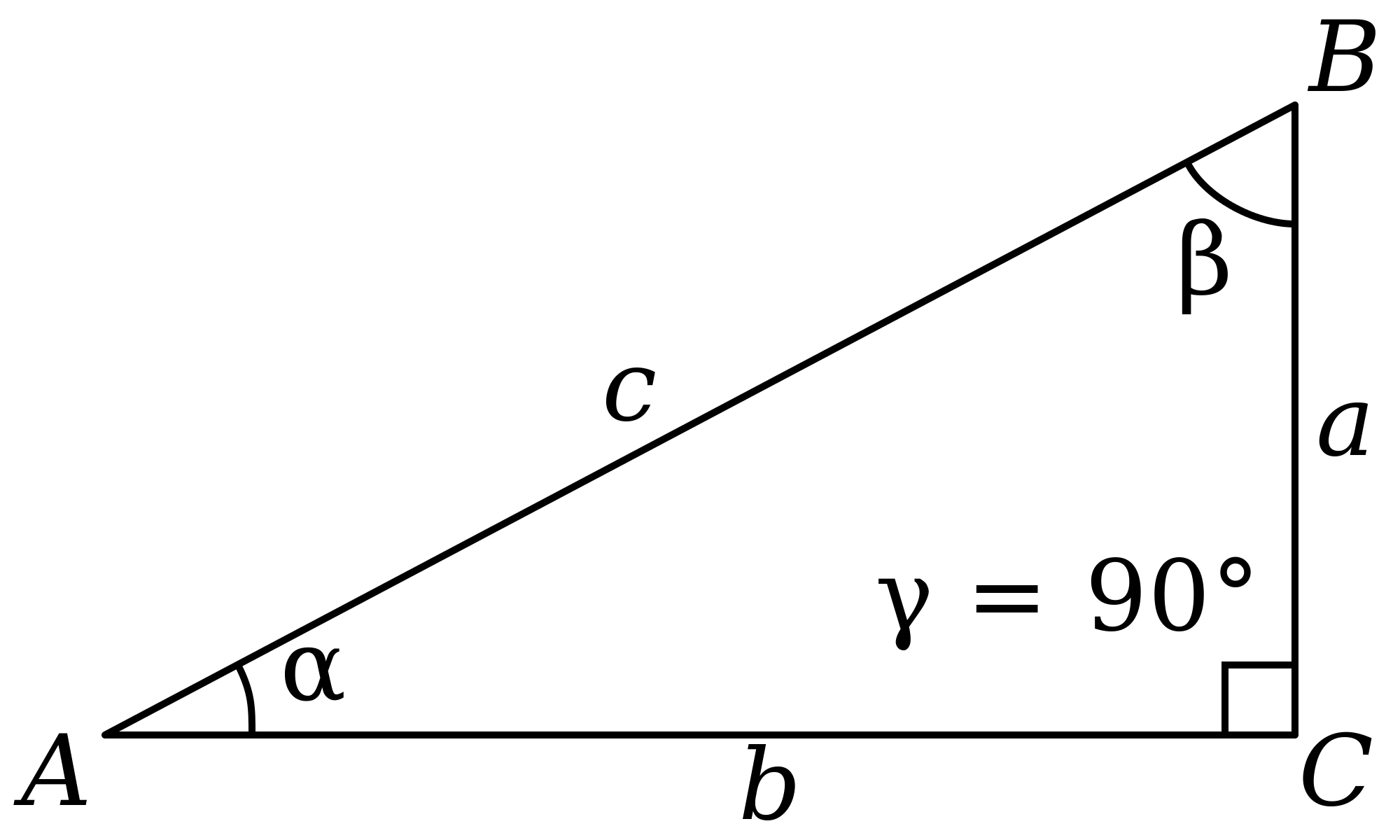}}
 \caption{A right-angled triangle with $\gamma$ being the right angle (Dmitry Fomin, \url{https://commons.wikimedia.org/wiki/File:Right_triangle_with_notations_(black).svg},CC0).}
   \label{f:triangle}
\end{figure}

The hypotenuse is the side of a triangle opposite of the right angle. In Fig.~\ref{f:triangle}, it is c. The other sides are called legs or catheti. The leg opposite to a given angle is called the opposite leg, while the other is the adjacent leg. In a right-angled triangle, the relations between the legs and the hypotenuse are expressed as trigonometric functions of the angles.
\begin{eqnarray}
 \sin\alpha &=& \frac{a}{c} = \frac{\rm opposing\ leg}{\rm hypotenuse}\\
 \cos\alpha &=& \frac{b}{c} = \frac{\rm adjacent\ leg}{\rm hypotenuse}\\
 \tan\alpha &=& \frac{a}{b} = \frac{\rm opposing\ leg}{\rm adjacent leg}
\end{eqnarray}
The Pythagorean Theorem tells us something about the relations between the three legs of a right-angled triangle. It is named after the ancient Greek mathematician Pythagoras and says that the sum of the squares of the catheti is equal to the square of the hypotenuse.
\begin{equation}
c^2 = a^2 + b^2
\end{equation}
For general triangles, this expands to the law of cosines.
\begin{equation}
c^2 = a^2 + b^2 - 2ab\cdot \cos\gamma
\end{equation}
For $\gamma = 90\degr$ it reduces to the Pythagorean Theorem.

\subsection{Early navigational skill}
Early seafaring peoples often navigated along coastlines before sophisticated navigational skills were developed and tools were invented. Sailing directions helped to identify coastal landmarks \citep{hertel_geheimnis_1990}. To some extent, their knowledge about winds and currents helped them to cross short distances, like e.g. in the Mediterranean.

Soon, the navigators realised that celestial objects, especially stars, can be used to keep the course of a ship. Such skills have been mentioned in early literature like Homer’s Odyssey which is believed to date back to the 8th century BCE. There are accounts of the ancient people of the Phoenicians who were able to even leave the Mediterranean and ventured on voyages to the British coast and even several hundred miles south along the African coast \citep{johnson_history_2009}. A very notable and well documented long distance voyage has been passed on by ancient authors and scholars like Strabo, Pliny and Diodorus of Sicily. It is the voyage of Pytheas, a Greek astronomer, geographer and explorer from Marseille who around 300 BCE apparently left the Mediterranean by passing Gibraltar and made it up north until the British Isles and beyond the Arctic Circle, where he possibly reached Iceland or the Faroe Islands that he called Thule \citep{baker_ancient_1997}. Pytheas already used a gnomon or a sundial, which allowed him to determine his latitude and measure the time during his voyage \citep{nansen_northern_1911}.

\subsection{Sailing along a latitude}
At these times, the technique of sailing along a parallel (of the equator) or latitude was used by observing
\newglossaryentry{circpol}
{
         name = {Circumpolar},
  description = {Property of celestial objects that never set below the horizon.}
}
circumpolar stars. The concept of latitudes in the sense of angular distances from the equator was probably not known. However, it was soon realised that when looking at the night sky, some stars within a certain radius around the celestial poles never set; they are circumpolar. When sailing north or south, sailors observe that the celestial pole changes, too, and with it the circumpolar radius. Therefore, whenever navigators see the same star
\newglossaryentry{culm}
{
         name = {Culmination},
  description = {Passing the meridian of celestial objects. These objects attain their highest or lowest elevation there.}
}
culminating – transiting the meridian – at the same elevation, they stay on the ``latitude''.  For them, it was sufficient to realise the connection between the elevation of stars and their course. Navigators had navigational documents that listed seaports together with the elevation of known stars. In order to reach the port, they simply sailed north or south until they reached the corresponding latitude and then continued west or east.

Nowadays, the easiest way to determine one's own latitude on Earth is to measure the elevation of the North Star, Polaris, as a proxy for the true celestial North Pole. In our era, Polaris is less than a degree off. However, 1000 years ago, it was $8\degr$ away from the pole.

\subsection{The Kamal}
\begin{figure}[!ht]
 \resizebox{\hsize}{!}{\includegraphics{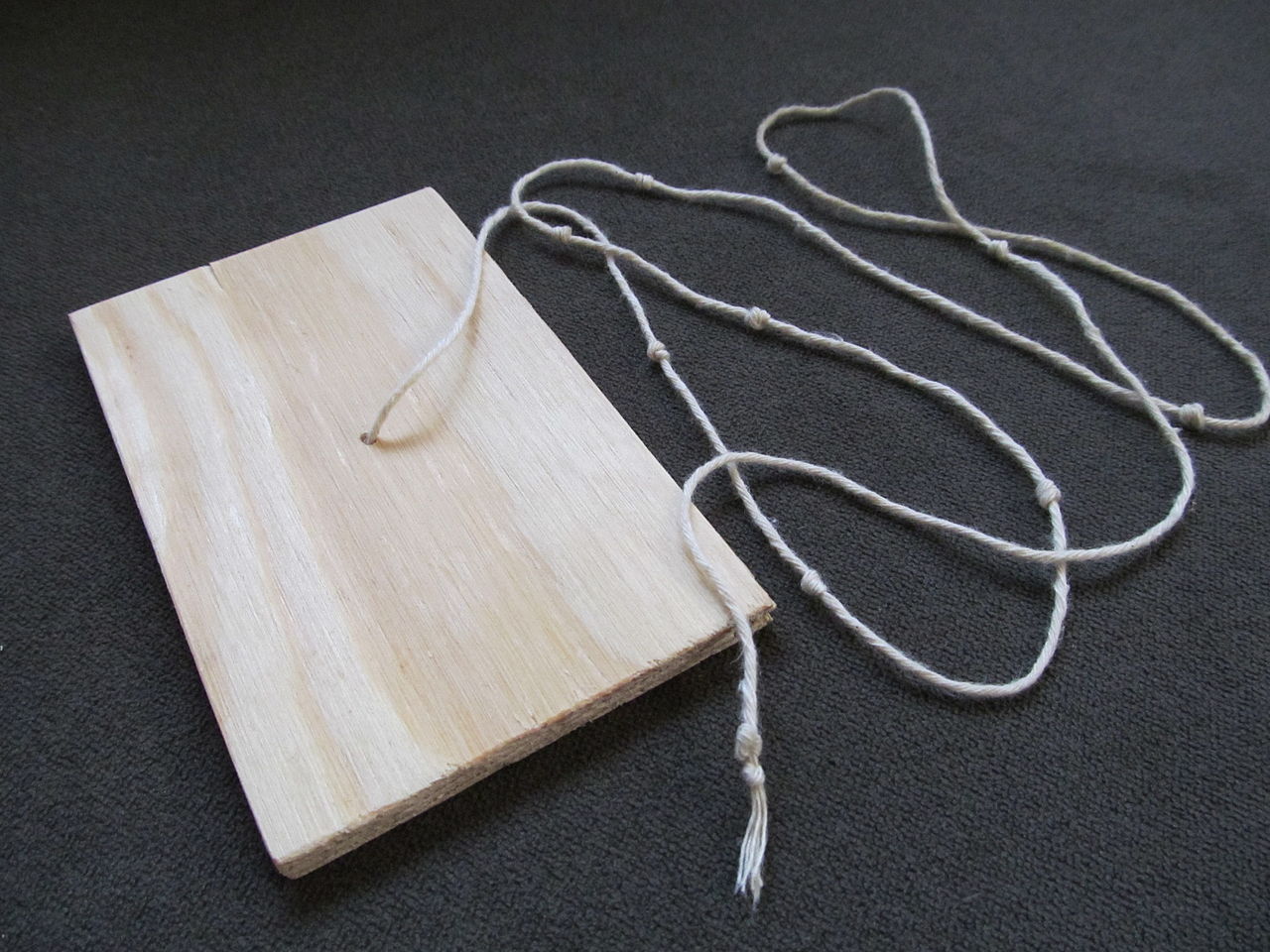}}
 \caption{A simple wooden kamal. It consists of a surveying board and a cord with a line of knots (Bordwall, \url{https://commons.wikimedia.org/wiki/File:Simple_Wooden_Kamal_(Navigation).jpg}, ``Simple Wooden Kamal (Navigation)'', \url{https://creativecommons.org/licenses/by-sa/3.0/legalcode}).}
   \label{f:kamal1}
\end{figure}

{\setcode{utf8}
The kamal (Fig.~\ref{f:kamal1}) is a navigational tool invented by Arabian sailors in the 9th century CE \citep{mcgrail_boats_2001}. Its purpose is to measure stellar elevations without the notion of angles. If you stretch out your arm, one finger subtends an angle. This method appears to have been the earliest technique to determine the elevation of stars. In the Arabian world, this ``height'' is called {\em isba} (\<إصبع >) which simply means {\em finger}. The corresponding angle is $1\degr 36'$ \citep{malhao_pereira_stellar_2003}.}

\begin{figure}[!ht]
 \resizebox{\hsize}{!}{\includegraphics{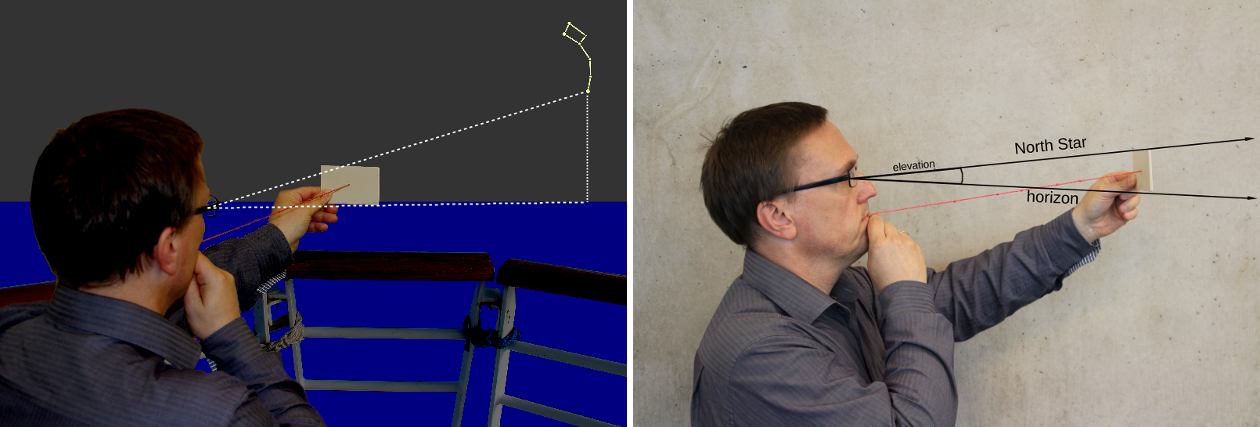}}
 \caption{Illustration of how the kamal was used to measure the elevation of a star, in this case Polaris. The lower edge was aligned with the horizon. Then, the distance between the eyes and the kamal was modified until the upper edge touched the star. The distance was set by knots tied into the cord that was held between the mouth and the kamal. The knots indicated elevations of stars (M. Nielbock,  \url{https://commons.wikimedia.org/wiki/File:Kamal_Polaris.png}, \url{https://commons.wikimedia.org/wiki/File:Kamal_Polaris_Side.png}, \url{https://creativecommons.org/licenses/by/4.0/legalcode}).}
   \label{f:kamal2}
\end{figure}

This method was standardised by using a wooden plate, originally sized roughly 5 cm x 2.5 cm, with a cord attached to its centre. When held at various distances, the kamal subtends different angles between the horizon and the stars (Fig.~\ref{f:kamal2}). Knots located at different positions along the cord denote the elevations of stars and, consequently, the latitude of various ports.

When Vasco da Gama set out to find the sea passage from Europe to India in 1497, he stopped at the Eastern African port of Melinde (now: Malindi), where the local Muslim Sheikh provided him with a skilled navigator of the Indian Ocean to guide him to the shores of India. This navigator used a kamal for finding the sailing directions \citep{launer_navigation_2009}.

Since the latitudes the Arabian sailors crossed during their passages through the Arabian and Indian Seas are rather small, the mentioned size of the kamal is sufficient. For higher latitudes, the board must be bigger to avoid very small lengths of the cord to realise such angles.

\begin{figure}[!ht]
 \resizebox{\hsize}{!}{\includegraphics{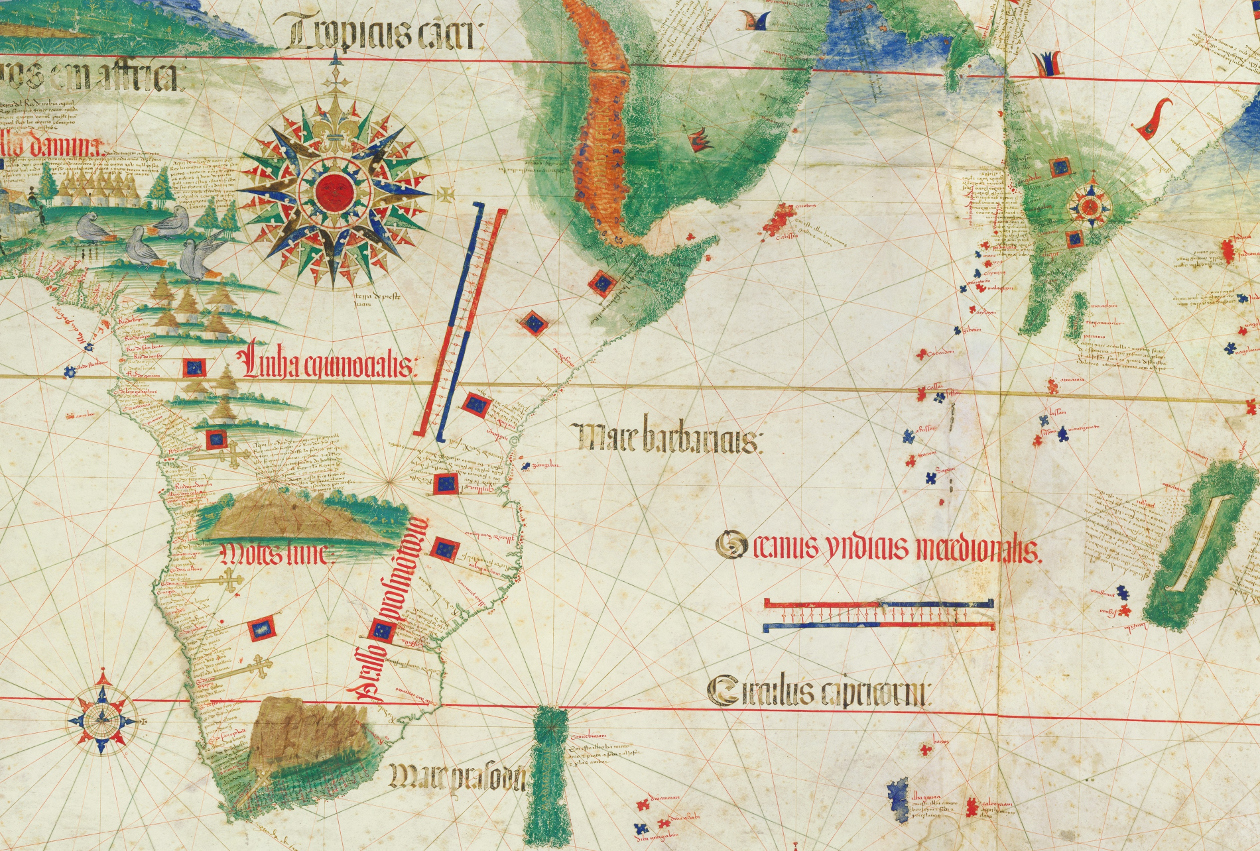}}
 \caption{Excerpt of a world map from 1502 showing the Indian Ocean. All sea routes from the Arabian Peninsula and India lie between the Tropic of Cancer and the Equator. The port of Melinde is indicated at the third flag from the top at the eastern African coast (Cantino Planisphere, 1502, Biblioteca Estense Universitaria, Modena, Italy,
 \url{https://commons.wikimedia.org/wiki/File:Cantino_planisphere_(1502).jpg, public domain}).}
   \label{f:cantino}
\end{figure}

\subsection{The geometry of the kamal}
\begin{figure}[!ht]
 \resizebox{\hsize}{!}{\includegraphics{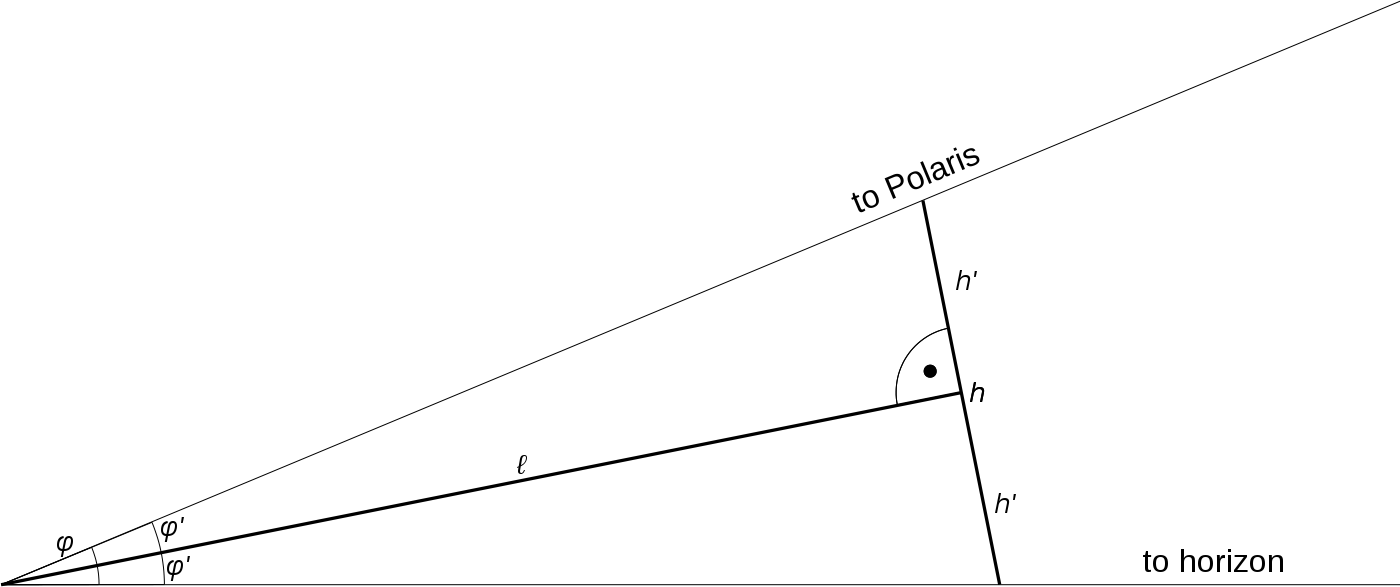}}
 \caption{Simplified geometry of the kamal which subtends an angle $\varphi$ between the horizon and Polaris. The kamal has a height labelled $h$. The length of the cord between the eyes and the kamal is labelled $\ell$ (M. Nielbock, own work).}
   \label{f:geo1}
\end{figure}

To measure an angle $\varphi$ with a kamal of height $h$ (Fig.~\ref{f:geo1}), the distance between the eyes and the board held perpendicularly to the line of sight needed is $\ell$. This is realised by a knot in the cord on the side opposite to the kamal board. In this simple configuration we get:
\begin{equation}
\ell = \frac{h'}{\tan\varphi} = \frac{h}{2\cdot \tan\left(\frac{\varphi}{2}\right)}
\end{equation}

\begin{figure}[!ht]
 \resizebox{\hsize}{!}{\includegraphics{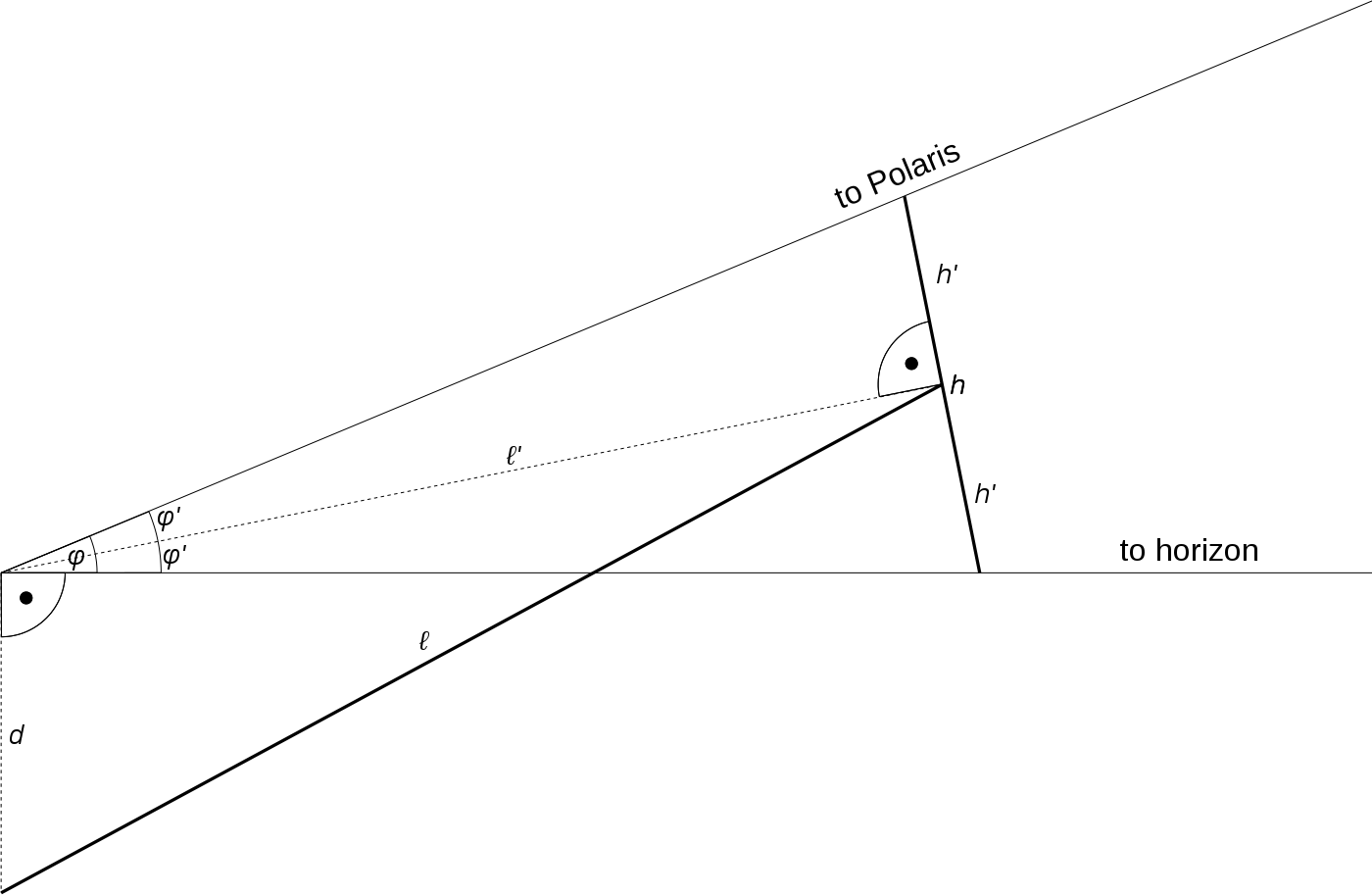}}
 \caption{More realistic geometry of the kamal considering the difference in distance between the kamal on one side and the mouth and the eyes on the other. The distance between the mouth and the eyes is labelled $d$ (M. Nielbock, own work).}
   \label{f:geo2}
\end{figure}

However, the measurement is done with the cord between the teeth or just in front of the lips. Eyes and mouth are separated by the length $d$ (Fig.~\ref{f:geo2}). The true length of the cord is then $\ell$, while $\ell'$ is the distance between the eyes and the kamal board that defines the angle $\varphi$. This more realistic approach leads to:
\begin{equation}
\ell = \sqrt{\frac{h}{2\cdot \tan\left(\frac{\varphi}{2}\right)}+d^2+hd\cdot\cos\left(\frac{\varphi}{2}\right)}
\label{e:kamal}
\end{equation}
We see that for $d=0$ we again get the simplified version above. The difference between $\ell$ and $\ell'$ can be a few centimetres. A realistic value is $d=7$\,cm.

This geometry is accurate enough for uncertainties inherent to the measuring method. Note that it is always assumed that the kamal board is held in an angle perpendicular to the line of sight, not the cord. In addition, the horizon is assumed to be the mathematical one (Fig.~\ref{f:poleheight}). This means that the dip of the visible horizon is neglected.

\section{List of material}
The list contains items needed by one student. Some of them can be shared by two to four individuals.
\begin{itemize}
 \item Worksheets
 \item Kamal or materials and instructions to build one
 \item Pencil
 \item Torch (for outside activity)
\end{itemize}

\noindent
Constructing the kamal:
\begin{itemize}
 \item one piece of ply wood (preferred) or very stiff card board (21 cm x 12 cm x 4 mm)
 \item 50 cm of cord
 \item Pencil
 \item Ruler
 \item Saw (for the wood) or scissors (for the cardboard), if the board has to be cut to fit the size needed
 \item Drill (for the wood) or thick needle (for the cardboard)
\end{itemize}

\section{Goals}
With this activity, the students will learn that
\begin{itemize}
 \item celestial navigation and corresponding tools have been developed already many centuries ago.
 \item the kamal is a simple tool to measure the elevations of stars.
 \item with the kamal we can nowadays easily determine our latitude on Earth.
\end{itemize}

\section{Learning objectives}
\begin{itemize}
 \item Students will build their own historical navigational instrument to understand how medieval mariners have used the stars to navigate.
 \item They will use it to determine their latitude on earth, and so understand, how rather simple and accurate such skills are.
 \item In the course of the activity, the students will learn to find Polaris, the North Star, to be able to determine the cardinal directions during night, which provides them with basic knowledge for navigating the seas.
\end{itemize}

\section{Target group details}

\noindent
Suggested age range: 14 -- 16 years\\
Suggested school level: Middle School, Secondary School\\
Duration: 90 minutes

\section{Evaluation}
\begin{itemize}
 \item There are detailed building instructions for the kamal included. It is very simple anyway. The result of the latitude measurement can be easily checked with online resources. This is also part of the activity.
 \item The teacher is responsible for providing a basic background on latitudes and longitudes. However, the success of learning can be judged with the questions and answers provided.
 \item Finding Polaris is a prerequisite for the success of this activity. If this activity is held as a group experience, the students can support each other. In addition, the teacher can guide the students by
 \begin{itemize}
  \item  presenting a planetarium software for practising.
  \item visiting a planetarium.
  \item using a laser pointer during the field experiment.
 \end{itemize}
\end{itemize}

\section{Full description of the activity}
\subsection{Introduction}
It would be beneficial, if the activity be included into a larger context of seafaring, e.g. in geography, history, literature, etc.

Tip: This activity could be combined with other forms of acquiring knowledge like giving oral presentations in history, literature or geography highlighting navigation. This would prepare the field in a much more interactive way than what a teacher can achieve by summarising the facts.

Tip: There are certainly good documentaries available on sea exploration and navigation that could be shown as an introduction.

\medskip\noindent
Episode 2: Celestial Navigation (Duration: 4:39)\\
\url{https://youtu.be/DoOuSo9qElI}

\medskip\noindent
How did early Sailors navigate the Oceans? | The Curious Engineer (Duration: 6:20)\\
\url{https://youtu.be/4DlNhbkPiYY}

\medskip\noindent
Isn't that India? - Navigation at Sea I PIRATES (Duration: 5:56)\\
\url{https://youtu.be/OCPnmfe5PJ4}

\medskip\noindent
Navigation in the Age of Exploration (Duration: 7:05)\\
\url{https://youtu.be/X3Egmp94aZw}

\medskip\noindent
World Explorers in 10 Minutes (Duration: 9:59)\\
\url{https://youtu.be/iUkOfzhvMMs}

\medskip\noindent
Once upon a time … man: The Explorers - The first navigators (Duration: 23:13)\\
\url{https://youtu.be/KuryXLnHsEY}

\medskip\noindent
The Ancient Seamasters (Duration: 1:29:07)\\
\url{https://youtu.be/47kAtmYTCmY}

\medskip
Ask the students, if they had an idea for how long mankind already uses ships to cross oceans. One may point out the spread of the Homo sapiens to islands and isolated continents like Australia.

\medskip\noindent
Possible answers:\\\noindent
We know for sure that ships have been used to cross large distances already since 3,000 BCE or earlier. However, the early settlers of Australia must have found a way to cross the Oceans around 50,000 BCE.

\medskip
Ask them, what could have been the benefit to try to explore the seas. Perhaps, someone knows historic cultures or peoples that were famous sailors. The teacher can support this with a few examples of ancient seafaring peoples, e.g. from the Mediterranean.

\medskip\noindent
Possible answers:\\\noindent
Finding new resources and food, trade, spirit of exploration, curiosity.

\medskip
Ask the students, how they find the way to school every day. What supports their orientation to not get lost? As soon as reference points (buildings, traffic lights, bus stops, etc.) have been mentioned ask the students, how navigators were able to find their way on the seas. In early times, they used sailing directions in connection to landmarks that can be recognised. But for this, the ships would have to stay close to the coast. Lighthouses improved the situation. Magnetic compasses have been a rather late invention around the 11th century CE, and they were not used in Europe before the 13th century. But what could be used as reference points at open sea? Probably the students will soon mention celestial objects like the Sun, the Moon and stars.

Tell the story of the kamal and Vasco da Gama, the discoverer of the direct passage from Europe to India. See corresponding section in the background material and:
\begin{itemize}
\item \url{https://archive.org/stream/vascodagamahisvo00towl#page/136/mode/2up}
\item \url{http://www.heritage-history.com/?c=read&author=towle&book=dagama&story=king}
\end{itemize}

\subsubsection{Activity 1: Building the kamal}
This can be done by the teacher prior to the activities or introduced as an additional exercise for the students. An instruction manual is available separately.

\noindent
Material needed:
\begin{itemize}
\item one piece of ply wood (preferred) or very stiff card board (21 cm x 12 cm x 4 mm)
\item 50 cm of cord
\item Pencil
\item Ruler
\item Saw (for the wood) or scissors (for the cardboard), if the board has to be cut to fit the size needed
\item drill (for the wood) or thick needle (for the cardboard)
\end{itemize}

The kamal was originally conceived as a navigational tool for low latitudes. Therefore, its size was relatively small, i.e. a few centimetres. This was enough to measure angles of $10\degr$ to $20\degr$ above the horizon. For example, for a kamal of 5~cm in height, a cord length of 20~cm yields an elevation measure of $15\degr$. However, this relation is not linear. Therefore, for higher latitudes a larger kamal board is needed. A good compromise is a height of 21~cm, while the width can be 12~cm. With these dimensions, the following relations hold. For very low latitudes, the kamal can be rotated by $90\degr$, so that the smaller width permits smaller cord lengths to reach the same angles.

\begin{table}
\centering
\caption{Dimensions and relations between angles and lengths of a kamal according to Eq.~\ref{e:kamal}. The distance between the eyes and the mouth is assumed to be 7~cm. The recommended combinations are given in bold face.}
\label{t:kamal}
\begin{tabular}{ccccc}
\hline\hline
Angle     & Board  & Cord   & Board & Cord \\
subtended & height & length & width & length \\
$(\degr)$ & (cm)   & (cm)   & (cm)  & (cm) \\
\hline
30 & 21 & 41.6 & \bf 12 & \bf 25.1 \\
35 & 21 & 36.0 & \bf 12 & \bf 22.2 \\
40 & 21 & 31.9 & \bf 12 & \bf 20.0 \\
45 & \bf 21 & \bf 28.8 & 12 & 18.3 \\
50 & \bf 21 & \bf 26.3 & 12 & 17.0 \\
55 & \bf 21 & \bf 24.2 & 12 & 16.0 \\
60 & \bf 21 & \bf 22.5 & 12 & 15.2 \\
65 & \bf 21 & \bf 21.1 & 12 & 14.4 \\
70 & \bf 21 & \bf 19.9 & 12 & 13.8 \\
\hline
\end{tabular}
\end{table}

For each kamal, prepare a thin piece of ply wood (approx.~4~mm) with 21~cm~x~12~cm in size. If that is not available, a piece of very stiff cardboard of equal size is also possible. Determine the centre of the board by drawing or scratching two diagonal lines that connect opposite corners. Drill a hole through the centre that is big enough to permit the cord to fit through. It must also be small enough to not let it slide out again after tying a knot.

\begin{figure}[!ht]
 \resizebox{\hsize}{!}{\includegraphics{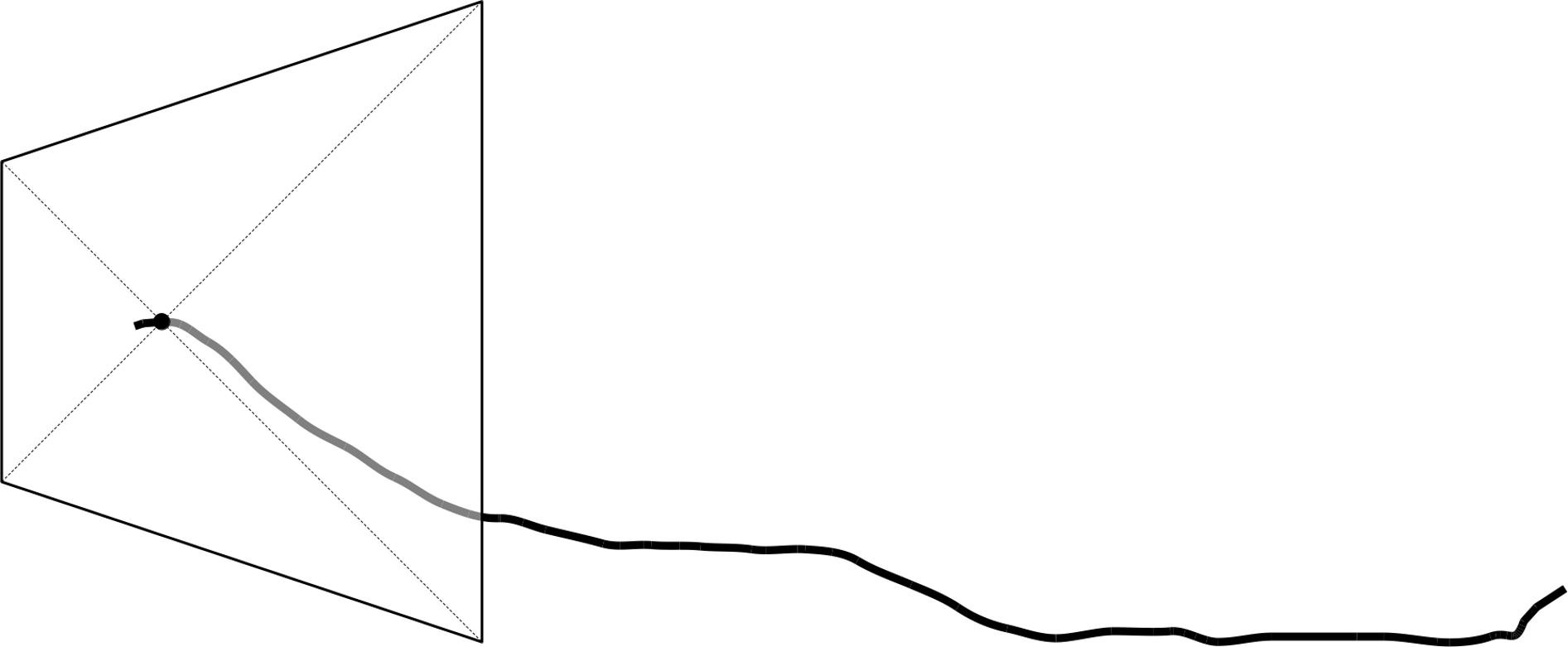}}
 \caption{The kamal after running the cord in the central hole (M. Nielbock, own work).}
   \label{f:template}
\end{figure}

Tie a knot at one end of the cord and run it in the central hole of the board. The knot should block the cord from sliding through the hole.

Now add knots at distances from the board as indicated in Tab.~\ref{t:kamal}. Be careful to keep the cord straightened. You can restrict the number of knots according to the angular range needed for the activities. Remember that the elevation of Polaris corresponds to the latitude.

Fill out the table on the worksheet that lists the number of knots and the corresponding angles.

\subsection{Activity 2: Angles in the sky}
\subsubsection{Introduction}
The worksheets contain Fig.~\ref{f:trails} (star trails). There are a few questions to be asked that can help understand the concept of the apparent trajectories of stars.

\medskip\noindent
Q: What does this picture show, in particular, where the bright curved lines come from?\\\noindent
A: As the Earth rotates, the stars seem to revolve around a common point. This is the celestial pole. Long exposures visualise the path of the stars as trails.

\medskip\noindent
Q: How does the picture show us that some stars do not set or rise during a full day?\\\noindent
A: Many trails can be followed to form a full circle. One rotation is 24 hours.

\medskip\noindent
Q: Do you know a star that is next to the celestial North Pole? In this picture, it should be close to the centre of rotation.\\\noindent
A: This is Polaris or the North Star. It is the star that produces the smallest trail close to the centre of the trails.

\medskip\noindent
Q: Imagine you are at the terrestrial North Pole. Where would Polaris be in the sky? Where would it be, if you stood at the equator?\\\noindent
A: North Pole: Zenith, i.e. directly above; Equator: at the northern horizon

\subsubsection{Preparations}
Find a spot outside with a good view to the northern sky and the horizon. This activity can be done as soon as the North Star is visible. Therefore, the summer time may not the best season to carry out this activity.

\subsubsection{Finding Polaris}
Finding Polaris in the sky is rather simple. As soon as the stars are visible, let the students look at them for a while and ask them, if they knew the group of stars that is often called the Big Dipper. Its name is different in different cultures (Ladle, Great Chariot, Plough, Drinking Gourd). It is easy to find in the northern hemisphere as it is always above the horizon. A video explains it in detail.

\medskip\noindent
Find North with the Stars -- Polaris \& Ursa Major - Celestial Navigation (Duration: 11:04)\\\noindent
\url{https://youtu.be/n_gT9nBfhfo}

\medskip
Figure~\ref{f:polaris} also shows how Polaris can be found using the Big Dipper (also available as individual image). It is contained in the worksheet. Find the box of the stellar group and the two stars to the front ($\alpha$, $\beta$). Extend the line between them five times and find a moderately bright star. This is Polaris, the North Star.

\subsubsection{Measuring the elevation of Polaris}
Now the students use the kamal. The cord must be kept straightened during measurements. The board must be held according to the suggestions given in Tab.~\ref{t:kamal} and perpendicular to the line of sight. Any tilt would compromise the measurement.

Similar as shown in Fig.~\ref{f:kamal2} (provided in the worksheet), the lower edge of the kamal must be aligned with the horizon. Then, the length of the cord is modified until the upper edge touches the star. The alignment with the horizon and the star should be checked again.

The students count the number of knots needed to keep the kamal aligned. Counting starts with the knot closest to the board. Perhaps, they will have to interpolate the position between two knots. They write down the number and read off the corresponding angle from the list in their worksheet. This is the latitude they will have determined.

The values of the various individuals and groups may differ. 

\medskip\noindent
Q: Why are the results not always identical?\\\noindent
A: Manufacturing not perfect (esp. knot positions), different sizes change perspective a bit, kamal was not always held correctly

\medskip\noindent
Q: How would this affect real navigation on open seas?\\\noindent
A: Small errors of a few degrees can lead to course deviations. One degree in latitude corresponds to 60 nautical miles. Repeated measurements and additional information can mitigate this effect.

\subsubsection{Analysis}
This can be done as homework and compared during the next lesson in school. Let the students check their result with a local map that provides coordinates or on-line services like Google Maps or Google Earth.

In Google Maps, you can right-click on your location and then click on “What's here?”. A small window appears at the bottom of the screen that lists two numbers. The first is the latitude in degrees with decimals. This number is added to the worksheet.

The students may realise that the result differs from their own measurement. Let them write down reasons, why that is the case. During the next lesson, let them discuss their results.

\section{Connection to school curriculum}
This activity is part of the Space Awareness category ``Navigation Through The Ages'' and related to the curricula topics:
\begin{itemize}
\item Coordinate systems
\item Basic concepts, latitude, longitude
\item Celestial navigation
\item Instruments
\end{itemize}

\section{Conclusion}
The kamal is a navigational tool that was invented by Arab navigators and has been used for many centuries since. This activity uses the example of the kamal to demonstrate how navigation at sea can be successful with some knowledge about astronomy and the stars combined with simple tools. The students learn some major aspects of the history of navigation by applying basics of math and astronomy. They build their own kamal and learn how to use it to determine their latitude on Earth by using Polaris as the resting reference point in the sky. With his activity, they get a feeling for what it took to find one's way on the oceans.

\begin{acknowledgements}
This resource was developed in the framework of Space Awareness. Space Awareness is funded by the European Commission’s Horizon 2020 Programme under grant agreement no. 638653.
\end{acknowledgements}

\bibliographystyle{aa}
\bibliography{Navigation}

\begin{thebibliography}{7}
\expandafter\ifx\csname natexlab\endcsname\relax\def\natexlab#1{#1}\fi

\bibitem[{Baker \& Baker(1997)}]{baker_ancient_1997}
Baker, R.~F. \& Baker, C.~F. 1997, Ancient {Greeks}: {Creating} the {Classical}
  {Tradition} (New York, USA: Oxford University Press)

\bibitem[{Hertel(1990)}]{hertel_geheimnis_1990}
Hertel, P. 1990, Das {Geheimnis} der alten {Seefahrer}, Geographische
  {Bausteine} No.~38 (Gotha, Germany: Hermann Haack Verlagsgesellschaft mbH)

\bibitem[{Johnson \& Nurminen(2009)}]{johnson_history_2009}
Johnson, D.~S. \& Nurminen, J. 2009, The {History} of {Seafaring}, 2nd edn.
  (National Geographic), authorised German Edition

\bibitem[{Launer(2009)}]{launer_navigation_2009}
Launer, D. 2009, Navigation {Through} {The} {Ages}, ed. J.~Mc~Geary \& J.~Simon
  (Dobbs Ferry, NY 10522: Sheridan House Inc.)

\bibitem[{Malhão~Pereira(2003)}]{malhao_pereira_stellar_2003}
Malhão~Pereira, J.~M. 2003, Proceedings of the International Seminar on Marine
  Archeology

\bibitem[{McGrail(2001)}]{mcgrail_boats_2001}
McGrail, S. 2001, Boats of the {World}: {From} the {Stone} {Age} to {Medieval}
  {Times} (Oxford, UK: Oxford University Press)

\bibitem[{Nansen(1911)}]{nansen_northern_1911}
Nansen, F. 1911, In {Northern} {Mists}: {Arctic} {Exploration} in {Early}
  {Times}, Vol.~1 (New York: Frederick A. Stokes company)

\end{thebibliography}

\glsaddall
\printglossaries

\begin{appendix}
\section{Connection to other educational materials}
This unit is part of a larger educational package called “Navigation Through the Ages” that introduces several historical and modern techniques used for navigation. An overview is provided via:

\href{http://www.space-awareness.org/media/activities/attach/b3cd8f59-6503-43b3-a9e4-440bf7abf70f/Navigation\%20through\%20the\%20ages\%20compl\_z6wSkvW.pdf}{Navigation\_through\_the\_Ages.pdf}

 \section{Supplemental material}
 The supplemental material is available on-line via the Space Awareness project website at \url{http://www.space-awareness.org}. The direct download links are provided as follows:
 
 \begin{itemize}
  \item Worksheets: \href{https://drive.google.com/file/d/0Bzo1-KZyHftXOGIyUnFkTS1SMGM/view?usp=sharing}{astroedu1647-The-Kamal-WS.pdf}
\end{itemize}
\end{appendix}
\end{document}